\documentstyle[preprint,aps]{revtex}
\newcommand{\half}{\frac{1}{2}}

\newcommand{\bk}{{\bf{k}}}
\begin{document}

\title{ Transition temperature and a spatial dependence of the
superconducting gap for multilayer high-temperature superconductors}

\author{Krzysztof Byczuk\footnote{e-mail: byczuk@fuw.edu.pl;
 jozef@physics.purdue.edu }}

\address{Institute of Theoretical Physics, Warsaw University,
ulica Ho\.za 69, 00-681 Warszawa, Poland}

\author{Jozef Spa\l ek*}
\address{Institute of Physics, Jagiellonian University,
ulica Reymonta 4, 30-059 Krak\.ow, Poland;\\
Department of Physics, Purdue University, West Lafayette, IN 47907-1396}

\date{\today}
\maketitle
\date{August 25, 1995}
\vspace*{24pt}

%\begin{abstract}
We derive the expressions for the transition temperature $(T_{c})$, and the
spatial dependence of the superconducting gap for a multilayer high-$T_{c}$
superconductor composed of groups of tightly spaced planes separated by a
larger distance.  The results are compared with experiment and provide a
strong support for an interlayer hopping as the driving force of the large
$T_{c}$ enhancement in multilayered compounds.  Our results are universal
in the sense that they are valid for an arbitrary pairing potential $V_{\bf
k k'}$ in the $\rm CuO_{2}$ planes, as well as for both Fermi and nonFermi
liquids.
%\end{abstract}
\vspace*{12pt}

\newpage

The physical origin of pairing in high temperature superconductors has
not been determined as yet.  Therefore, it is very important to
formulate a language describing those systems, which is independent of
the detailed underlying electronic structure, but still concrete enough
to allow for an extensive experimental test.  We believe that the model
proposed by Anderson and his coworkers [1] and based on the pairing
enhancement due to an interlayer Cooper-pair tunneling provides such a
crucial feature, on the basis of which one can construct an appropriate
quantitative language.  Universal features of this model not discussed
so far are detailed below and compared with experiment.

In this paper we derive analytically the space profile of the
superconducting gap parameter assuming only that we have a
$\bk$-dependent pairing potential $V_{\bk\bk'}$ in the two spatial
dimensions (for a square planar configuration), and a coherent
(Josephson) pair tunneling between the planes [1,2].  Moreover, we
analyze explicitly the periodic arrangement of groups of planes, each
of them containing $p$ tightly spaced identical $\rm CuO_{2}$ planes,
and separated by a larger distance, but in contact with each other also
by, albeit weaker, Cooper pair tunneling.  Such a configuration is
drawn schematically in Fig.\ 1a for  the high temperature
superconductor.  The system is thus characterized by the intragroup
$(T^{(l)} (\bk))$ and intergroup $(\widetilde{T}(\bk))$ tunneling
amplitudes.  Hence, our approach can be applied to both mono- and
multi-layer high-$T_{c}$ systems and thus generalizes previous
treatments [1,3,4].  The calculated below space profile of the
superconducting gap is shown schematically in Fig.\ 1b.  Our results
for the transition temperature $T_{c}$ for either Fermi, or Luttinger
or statistical spin liquid, the three universality classes appearing
for the interacting fermion systems [5].

We label the single electron states by  the in-plane component of the wave
vector $\bk \equiv (k_{x},k_{y})$ and by spin quantum number $\sigma = \pm
1$, as well as by the two discrete numbers $(s,j)$, where $j = 1, \ldots,p$
characterizes the plane position within the group of tightly spaced planes,
and $s = 1, \ldots,N$ labels the groups.  The simplest Hamiltonian describing
such a nontrivial structure can be written as follows:
\begin{eqnarray}
H &=&
\sum_{s=1}^{N}\sum_{\bk \sigma}\sum_{j=1}^{p}
\epsilon_{\bk}b_{j,\bk \sigma}^{+s}
b_{j,\bk \sigma}^{s} -
\frac{1}{N}
\sum_{s=1}^{N}\sum_{\bk}\sum_{j=1}^{p}
V_{\bk\bk'}
b_{j,\bk\uparrow}^{+s}
b_{j,-{\bk}\downarrow}^{+s}
b_{j,-\bk '\downarrow}^{s}
b_{j,\bk'\uparrow}^{s}\nonumber \\
&-&
\sum_{s=1}^{N}\sum_{\bk}\sum_{j=1}^{p}\sum_{l=1}^{j+l\leq p}
T^{(l)}
( \bk)
\left[
b_{j,\bk \uparrow}^{+s}b_{j,-\bk\downarrow}^{+s}
b_{j+l,-\bk\downarrow}b_{j+l,\bk \uparrow}^{s} + H.C.
\right] \nonumber \\
&-&
\sum_{s=1}^{N}\sum_{\bk}\sum_{j=1}^{p}\sum_{l=1}^{1\leq j - l \leq p}
T^{(l)}
( \bk)
\left[
b_{j,\bk \uparrow}^{+s}b_{j,-\bk\downarrow}^{s}
b_{j-l,-\bk\downarrow}b_{j-l,\bk \uparrow}^{s} + H.C.
\right] \nonumber\\
&-&
\sum_{s=1}^{N-1}\sum_{\bk}
\widetilde{T} (\bk)
\left[
\left(
\sum_{j=1}^{p}
b_{j\bk \uparrow}^{+s}
b_{j,-\bk\downarrow}^{+s}
\right)
\left(
\sum_{j=1}^{p}
b_{j-\bk \downarrow}^{s+1}
b_{j,-\bk\uparrow}^{s+1}
\right)
+ H.C.
\right] .
\end{eqnarray}
The first two terms describe the BCS-type Hamiltonian for $N \cdot p$
independent planes, the second and the third express the pair tunneling
(with the matrix element $T^{(l)}(\bk)$) between the planes in the same group,
separated by $(l-1)$ superconducting planes in between, whereas the last term
represents
the tunneling between neighboring groups. We have assumed that the
intergroup tunneling matrix element $\widetilde{T}(\bk)$ is independent of
the pair of planes selected in those two groups, since the detailed balance
within the group must be achieved much faster than between the groups.

In the following we make the mean field (BCS-like) approximation, {\it
i.e.},  assume that the superconducting gap is characterized by the
nonvanishing average
$d_{j,\bk} \equiv < b_{j,\bk\uparrow}^{s}
b_{j,-\bk \downarrow}^{s} >$,
which will not depend on group index $s$, as each group is in identical
environment
and $N \rightarrow \infty$ (this, however, is not the case for the individual
planes within the group!)
In effect, the gap function can be defined as follows
\begin{eqnarray}
\Delta_{j,\bk} =
\sum_{j=1}^{p}
2\widetilde{T} (\bk)d_{j,\bk} +
\frac{1}{N}
\sum_{\bk ' }
v_{\bk\bk'}
d_{j, \bk'} +
\sum_{l=1}^{j+l \leq  p}
T^{(l)}
(\bk) d_{j+l,\bk} +
\sum_{l=1}^{1\leq j -l \leq p}
T^{(l)}
(\bk)
d_{j-1,\bk} .
\end{eqnarray}
The selfconsistency condition for the mean-field solution requires that
\begin{eqnarray}
d_{j,\bk} =
\Delta_{j,\bk}
\chi_{j,\bk} .
\end{eqnarray}
The generalized susceptibility $\chi_{j,\bk}$ in this universal relation
depends on the nature of the normal state.  In the case of the Fermi liquid
(FL) it takes the usual form [3].
\begin{eqnarray}
\chi_{j,\bk} =
\frac{1}{2E_{j,\bk}}
\tanh
\left(
\frac{\beta E_{j,\bk}}
     {2}
\right) ,
\end{eqnarray}
where the quasiparticle energy is
$E_{j\bk} =
\sqrt{(\epsilon_{\bk} - \mu)^{2} + | \Delta_{j,\bk} |^{2}}$.
In the case of the spin-charge separated Luttinger liquid (LL), {\it i.e.},
if we have the linear dispersion relation
$\epsilon_{\bk} - \mu \sim (k - k_{F})$
for bare particles and two Fermi velocities corresponding to charge $(v_{c})$
and spin $(v_{s})$ degrees of freedom, the function $\chi_{j,\bk}$ is [4]
\begin{eqnarray}
\chi_{j,\bk} =
\frac{|\Delta_{j,\bk}|^{2}}
     {\sqrt{\nu_{-}^{2} + 4 | \Delta_{j,\bk}|^{4}}}
\frac{\tanh(\beta E_{j,\bk}/2)}
     {E_{j,\bk}}
+
\frac{1}{\pi}
\int_{-\pi}^{\pi}
dx \tanh
\left(
\frac{\beta x}{2}
\right)
\frac{h_{s}h_{c}}
     {h_{s}^{2}h_{c}^{2} + | \Delta_{j,\bk}|^{4}}
\end{eqnarray}
where
$E_{j,\bk} = \half
[ \nu_{+} +
\sqrt{\nu_{-}^{2} + 4 | \Delta_{j,\bk}|^{4}}
] ,
h_{c,s} \equiv
\sqrt{|x^{2} - (v_{c,s}k)^{2}|},$
and
$\nu_{\pm} \equiv (v_{c}^{2} \pm v_{s}^{2}) k^{2}$.
Finally, for the  strongly correlated liquids with  non-standard
anticommutation relations for the fermion operators [5] the mean-field
approximation of the BCS-type leads also to Eq. (2), in which rhs is
renormalized only  by the factor
$A_{\bk} \equiv 1 - \bar{n}_{\bk}/2$, which near the Fermi energy has the
value $\half$.

We take into account only the coupling between the nearest neighbors, {\it
i.e.}, assume that $T^{(l)}(\bk) = T(\bk)\delta_{l,l\pm1}$.  This
approximation means that we can seek a solution in a simple exponential or
trigonometric form. In that situation the gap $\Delta_{j,\bk}$ is
determined by the system of universal difference equations
\begin{eqnarray}
\Delta_{1\bk} &-&
\frac{1}{N}
\sum_{\bk ' }
V_{\bk\bk'}
\chi_{1,\bk '}
\Delta_{1,\bk '} =
T(\bk)
\chi_{2, \bk}
\Delta_{2,\bk} +
2 \widetilde{T} (\bk)
\sum_{j ' = 1}^{p}
\chi_{j',\bk}
\Delta_{j',\bk} , \nonumber \\
\Delta_{j,\bk} &-&
\frac{1}{N}
\sum_{\bk '}
V_{\bk\bk '}
\chi_{j,\bk '}
\Delta_{j,\bk '} =
T(\bk)
(\chi_{j+1\bk}
\Delta_{j+1, \bk} +
\chi_{j-1,\bk}
\Delta_{j-1,\bk}) +
2\widetilde{T}
(\bk)
\sum_{j'=1}^{p}
\chi_{j',\bk}
\Delta_{j',\bk} , \nonumber \\
\Delta_{p,\bk} &-&
\frac{1}{N}
\sum_{\bk '}
V_{\bk\bk '}
\chi_{p,\bk '}
\Delta_{p,\bk '} =
T(\bk)
\chi_{p-1,\bk}
\Delta_{p-1,\bk} +
2\widetilde{T}
(\bk)
\sum_{j'=1}^{p}
\chi_{j',\bk}
\Delta_{j',\bk} .
\end{eqnarray}
The shape of the spatial profile of $\Delta _{j,\bk}$ ({\it i.e.}, its
dependence of $j$) should not depend on the system temperature, but only on
the form of the boundary conditions.  Therefore, we solve this system of
equations close to the transition temperature $T_{c}$, where the
generalized susceptibility $\chi_{j,\bk}$ can be linearized in the
following way
\begin{eqnarray}
\chi_{j,\bk}
\simeq
\frac{\tanh \left(\frac{\beta}{2}\epsilon_{\bk}\right)}
     {2\epsilon_{\bk}}
+
O (\Delta_{j,\bk}^{2}) ,
\end{eqnarray}
for FL, and
\begin{eqnarray}
\chi_{j\bk}
\simeq
\frac{1}{\pi}
\int_{v_{s}k}^{v_{c}k}
dx \tanh
\left(
\frac{\beta x}{2}
\right)
\frac{1}{h_{s}h_{c}} +
O(\Delta_{j,k}^{4}) ,
\end{eqnarray}
for LL.  To zero order in the gap parameter the generalized
susceptibilities do not depend on the layer index $j$.  With this
expansion, the system (6) represents a system of linear and  inhomogeneous
difference equations with respect to index $j$.  In order to solve them we
introduce two additional fictitious layers labeled by $j = 0$ and $j = p +
1$ [6].  The natural boundary conditions are expressed in terms of the
vanishing gap on these limiting layers, {\it i.e.}, we set
\begin{eqnarray}
\Delta_{0,\bk} =
\Delta_{p+1,\bk} = 0 .
\end{eqnarray}
These boundary conditions express the physical equivalence of all $p$
layers within the group, {\it i.e.}, we do not introduce the surface
layers, which differ from those inside.  In such a situation, the system of
equations (6) can be rewritten in a more symmetric form, both for the FL
and the LL cases, namely
\begin{eqnarray}
\Delta_{j,\bk} -
\frac{1}{N}
\sum_{\bk '}
V_{\bk\bk'}
\chi_{\bk '}
\Delta_{j,\bk '} =
T(\bk)
\chi_{\bk}
(\Delta_{j+1,\bk} +
\Delta_{j-1,\bk}) +
Z_{\bk} ,
\end{eqnarray}
where $Z_{\bk} \equiv 2\widetilde{T}(\bk)
\chi_{\bk}
\sum_{j ' = 1}^{p}
\Delta_{j',\bk}$,
and $j = 1\ldots p$.

We solve first the homogeneous part of Eq.\ (10) ({\it i.e.}, put $Z_{\bk}
\equiv 0$).  The spatial dependence of the solution for the gap can be taken in
the form
$\Delta_{j,\bk} = \Delta_{\bk}^{(\pm)} e^{\pm i\alpha j}$,
where the constants $\Delta_{\bk}^{(\pm)}$ fulfill the set of
self-consistent equations
\begin{eqnarray}
\Delta_{\bk}^{(\pm)}
(1- 2T
(\bk)
\chi_{\bk}
\cos \alpha) =
\frac{1}{N}
\sum_{\bk '}
V_{\bk\bk '}
\chi_{\bk '}
\Delta_{\bk '}^{(\pm)} .
\end{eqnarray}
Thus, the general solution of the homogeneous part is
\begin{eqnarray}
\Delta_{j,\bk} =
\Delta_{\bk}^{(+)}
e^{i\alpha j} +
\Delta_{\bk}^{(-)}
e^{-i\alpha j} ,
\end{eqnarray}
where the wave vector of the oscillating gap is determined from the
boundary conditions (1), and takes the form
\begin{eqnarray}
\left(
  \begin{array}{c c}
   1 & 1 \\
   e^{i\alpha(p+1)} &
   e^{-i\alpha(p+1)}
  \end{array}
\right)
\left(
  \begin{array}{c }
  \Delta_{\bk}^{(+)} \\
  \Delta_{\bk}^{(-)} \\
  \end{array}
\right)
= 0 .
\end{eqnarray}
The vanishing determinant of the above matrix provides a nontrivial
solution only when $\alpha = n \pi / (p + 1)$, where $n$ is an integer.
Substituting back the value of $\alpha$ to (13) we find $\Delta_{\bk}^{(+)}
= - \Delta_{\bk}^{(-)} \equiv \Delta_{\bk}$; so the solution (12) takes the
explicit form
\begin{eqnarray}
\Delta_{j,\bk} =
2i\Delta_{\bk}
\sin
\left(
\frac{n\pi j}
     {p + 1}
\right) .
\end{eqnarray}
Physically, the admissible solution is that with $n = 1$, as the solution
with no nodes inside $p$ layers should have the lowest energy.  The shape
of the spatial dependence of the gap has been drawn already
in Fig.\ 1b.  Finally, the gap magnitude $\Delta_{\bk}$ is determined from
Eq.\ (10) with $Z_{\bk} = 0$, which reduces to the integral equation:
\begin{eqnarray}
\Delta_{\bk} =
\frac{1}
     {1-2T(\bk)\chi_{\bk} \cos \left(\frac{\pi}{p+1}\right)}
\frac{1}{N}
\sum_{\bk}
V_{\bk\bk'}
\chi_{\bk '}
\Delta_{\bk '}
\end{eqnarray}

The solution  of the full Eq.\ (10) is found by superposing the general
solution(14) of the homogeneous part of (11) with a particular solution of the
full equation, {\it i.e.},  by taking
\begin{eqnarray}
\Delta_{j,\bk} =
\Delta_{\bk}^{'} +
\Delta_{\bk}^{(0)}
\sin
\left( \frac{\pi}{p+1} j\right) ,
\end{eqnarray}
where $\Delta_{\bk}^{(0)} = 2i\Delta_{\bk}$, and $\Delta_{\bk}^{'}$ is treated
as a small perturbation.  The part $\Delta_{\bk}^{'}$ does not depend on $j$
because the periodic boundary conditions can be taken for $N$ groups of
layers.  Substituting the solution (16) into (10) and omitting higher-order
terms ({\it i.e.}, those $\sim \Delta_{\bk}^{'} \cdot \widetilde{T}(\bk)$), we
have the  equation for $\Delta_{\bk}^{(0)}$
\begin{eqnarray}
\Delta_{\bk}^{(0)} =
\frac{1}
     {1-2[\widetilde{T}({\bk}) f (p) +
         T(\bk) \cos \left(\frac{\pi}{p+1}\right)]\chi_{\bk}}
\frac{1}{N}
\sum_{\bk\bk '}
V_{\bk\bk '}
\chi_{\bk '}
\Delta_{\bk '}^{(0)} ,
\end{eqnarray}
The equation for $\Delta_{\bk}^{'}$ is
\begin{eqnarray}
\Delta_{\bk}^{'} =
\frac{1}
      {1-2T(\bk)\chi_{\bk}}
\frac{1}{N}
\sum_{\bk '}
V_{\bk\bk '}
\chi_{\bk '}
\Delta_{\bk '}^{'}
\end{eqnarray}
where $f(p) \equiv \sin (\pi p/2(p+1))/\sin (\pi/2(p+1))$.

The physical $T_{c}$ is provided by the larger transition temperatures
determined from Eqs.\ (17) and (18).  Hence, it is always determined
from Eq.\ (17).  This feature of the theory illustrates the importance
of the intergroup hopping $\widetilde{T}(\bk)$ and formally appends the
ideas of [1].  Also, it reduces to that of Chakravarty {\it et al.} [1]
in the limit of $\widetilde{T} = 0$ and  $p=1$.  Most importantly, our
result is the same for both Fermi and Luttinger liquids, as well as for
the liquids with a generalized exclusion statistics described in
[5,7].

To discuss the implications of our approach let us consider first the
monolayer case $(p=1)$, {\it i.e.}, a material with the interlayer hopping
$\widetilde{T} (\bk) \equiv (t ')^{2}/t$, for which $t$ is the magnitude of
the bare single-particle hopping in the plane.  In that case a simple estimate
of the denominator of Eq.\ (17) is:
\begin{eqnarray}
1 - \widetilde{T} (\bk) \chi_{\bk}
\sim 1 - r
\frac{t'}{t} \cdot
\frac{t'}{k_{B}T_{c}} ,
\end{eqnarray}
where $r$ is a numerical constant of the order of unity.  Hence, the tunneling
can strongly enhance the pairing potential, since
$k_{B}T_{c} \sim r (t')^{2}/t$ [1].

To calculate explicitly the value of $T_{c}$ we consider explicitly the $p
\geq 1$ case with a constant pairing potential
$V_{\bk\bk'} = V$ in the energy regime of the width $\omega_{D} \ll \mu$.
Introducing the density of states $N(0)$ at the Fermi level we find from (17)
\begin{eqnarray}
1 =
VN(0)
\int_{-\omega_{D}}^{\omega_{D}}
d\epsilon
\frac{\chi(\epsilon)}
     {1-2[\tilde{\alpha} f(p) + \cos (\frac{\pi}{p+1})]
      \chi(\epsilon) (t')^{2}/t },
\end{eqnarray}
where $\tilde{\alpha}$ is the ratio of intergroup to intragroup tunneling
amplitudes.  In the extreme situation, when the interplanar tunneling provides
a dominant contribution to the effective pairing, we can make the assumption
$\omega_{D}/k_{B}T_{c} \ll 1$ and thus approximate $\tanh (x) \approx x$.
Here, we provide the $T_{c}$ derivation for the Luttinger liquid [4] using the
expression (8) for $\chi_{\bk}$.  In that case, Eq.\ (20) reads
\begin{eqnarray}
1 = V N (0)
\int_{0}^{\beta \omega_{D}/2} dx
\frac{I(x)}
     {1 - \beta T(p)I(x)} ,
\end{eqnarray}
where $T(p) \equiv
[\tilde{\alpha} f (p) +
\cos \left(\frac{\pi}{p+1}\right) ]
(t')^{2}/t$, and
\begin{eqnarray*}
I(x) =
\frac{1}{\pi x}
\int_{v_{s}/v_{c}}^{1} du
\frac{\tanh (xu)}
     {\sqrt{u^{2} - (u_{s}/u_{c})^{2}} \sqrt{1-u^{2}}} .
\end{eqnarray*}
In the limit $\beta \omega_{D} \ll 1$ the last integral is of elliptic form,
which has a property $\int_{a}^{b} dz/(\sqrt{z-a}\sqrt{b-z}) = \pi$; hence
$I(x) = 1$.  In effect, we obtain a remarkably simple expression for $T_{c}$
\begin{eqnarray}
T_{c} &=&
VN(0)
\frac{\omega_{D}}{2} +
\frac{1}{2}
\frac{(t')^{2}}{t}
\left[
\cos \left(
\frac{\pi}{p+1}
\right) +
\tilde{\alpha} f (p) \right]
\equiv T_{c}^{0} (p) +
T_{c}^{'}
\cos
\left(
\frac{\pi}{p+1}
\right) .
\end{eqnarray}
This result is of the same type as the formula derived for
$\tilde{\alpha} = 0$ in the Fermi liquid case [1] (it can be extended to the
case $\tilde{\alpha} \neq 0$ in straightforward manner).  Hence, the
transition temperature expression is universal and contains a combined
effect of the intrinsic in-plane pairing and the two interplanar
Josephson tunneling process.es

To test the relative role of the two tunneling contributions we have
fitted the tabulated [8] $T_{c}$ values to the expression (22).  Table
I displays the detailed comparison, together with the $T_{c}^{0}$ and
$T_{c}^{'}$ values.  The agreement between the theory and experiment is
excellent for $p > 1$ systems.  For $p=1$ systems we have chosen
$T_{c}^{0}$ as the middle point of the allowed interval; at least part
of the $T_{c}^{0}$ value must be ascribed to the interplanar tunneling
$\sim \tilde{\alpha} f(p)$.  In connection with this, one can say that
to have a proper behavior in the limit of large $p$ one must assume
that $\widetilde{T}({\bk}) \sim p^{-1}$; then $\tilde{\alpha} f (p)
\sim 1/p$, and the contribution due to the intergroup tunneling is
substantially smaller for larger $p$.  This is the reason why
$T_{c}^{0}$ value is essentially constant for $p > 1$.  From this point
of view, uncertainty in the $T_{c}$ values for the $p=1$ systems can be
understood by the planes intersection due to the sample imperfection
thus hampering the tunneling. Also, because of $T_{c}^{0}$ independence
on the systems with larger $p$, the coherence length along the axis
must exceed the smallest distance between the two neighboring groups.
Finally, the maximal available $T_{c}$ for each family is for $p
\rightarrow \infty$, and is also listed in Table I.

 From our results, one can draw a very important prediction: the $T_{c}$ should
increase substantially by applying the pressure along the $c$ axis, as the
pressure will increase it towards the value $T_{c}(\infty)$.

In summary, we have derived analytically the explicit forms of the
space profile of the superconducting gap, and the transition
temperature; these results were obtained without the need of specifying
the pairing potential in the plane, and independently of the nature of
the normal state.

The authors acknowledge the KBN Grants Nos. 2P302 093 05 and 2P302 171 06 in
Poland.  The work was performed, in part, at Purdue University(U.S.A.), where
it was supported by the MISCON Grant No. DE-FG-02-90ER 45427, and by the NSF
Grant No. INT. 93-08323.

\newpage
\begin{table}
\caption{Critial temperatures for various single and multilayer
superconductors}
\begin{tabular}{|l|c|c|c|c|c|}
Material & $p$ & $T_{c}^{\exp}$ & $T_{c}^{0}$ & $T_{c}^{'}$ & Theory: $T_{c}$
\\
\tableline
$\rm Bi_{2}Sr_{2}CuO_{6}$            & 1 & 0-20 & 10  & 148 & 10 \\
$\rm Bi_{2}Sr_{2}CaCuO_{8}$          & 2 & 85   & 10  & 148 & 84 \\
$\rm Bi_{2}Sr_{2}Ca_{2}Cu_{3}O_{10}$ & 3 & 110  & 10  & 148 & 112 \\
---                     & $\infty$  & -- & 10   & 148 & 158      \\
\tableline
$\rm Tl_{2}Ba_{2}CuO_{6}$ & 1 & 0--80 & 41 & 126 & 41 \\
$\rm Tl_{2}Ba_{2}CaCuO_{8}$ & 2 & 108 & 41 & 126 & 103 \\
$\rm Tl_{2}Ba_{2}Ca_{2}Cu_{3}O_{10}$ & 3 & 125 & 41 & 126 & 128 \\
--- & $\infty$ & -- & 41 & 126 & 167 \\
\tableline
$\rm Tl_{2}Ba_{2}CuO_{5}$ & 1 & 0--50 & 24 & 121 & 24 \\
$\rm Tl_{2}Ba_{2}CaCu_{2}O_{7}$ & 2 & 80 & 24 & 121 & 84 \\
$\rm Tl_{2}Ba_{2}Ca_{2}Cu_{3}O_{9}$ & 3 & 110 & 24 & 121 & 107 \\
$\rm Tl_{2}Ba_{2}Ca_{3}Cu_{4}O_{11}$ & 4 & 122 & 24 & 121 & 122 \\
--- & $\infty$ & -- & 24 & 121 & 145 \\
\tableline
$\rm HgBa_{2}CuO_{4}$ & 1 & 94 & 94 & 59 & 94 \\
$\rm HgBa_{2}Ca_{2}cu_{3}O_{8}$ & 3 & 135 & 94 & 59 & 135 \\
--- & $\infty$ & -- & 94 & 59 & 153
\end{tabular}
\end{table}
\newpage
\figure{Fig. 1. Schematic plot of the gap magnitude (top, a) for a multilayer
situation (bottom, b) involving groups of $p \geq 1$ tightly spaced planes.}
\end{document}